\documentclass[aps,prl,twocolumn,superscriptaddress]{revtex4-2}

\usepackage[hidelinks]{hyperref}
\hypersetup{
	colorlinks,
	linkcolor={red},
	citecolor={blue},
	urlcolor={blue}
}
\usepackage[hidelinks]{hyperref}

\usepackage{physics}
\usepackage{balance}
\usepackage{makecell}
\usepackage{suffix}
\usepackage{mathtools}
\usepackage[utf8]{inputenc}
\usepackage{booktabs}
\usepackage{cases}
\usepackage[multiple]{footmisc}
\usepackage{dcolumn}
\usepackage{color,soul}
\usepackage{rotating}
\usepackage{perpage}
\usepackage{siunitx}
\usepackage{xcolor}
\usepackage{soul}
\usepackage{amsmath}
\usepackage{tikz}
\usepackage[T1]{fontenc}
\usepackage{etoolbox}
\usepackage{graphics}
\usepackage{siunitx}
\usepackage{float}	
\usepackage{collref}
\usepackage{multirow}
\usepackage{mathtools}
\usepackage{bm}
\usepackage{url}
\usepackage{pifont}
\usepackage{balance}

\usepackage{amssymb}
\usepackage{mathtools}
\usepackage{amsmath}
\usepackage{bm}
\usepackage{siunitx}
\usepackage{graphicx}
\usepackage{float}
\usepackage{multirow}
\usepackage{booktabs}
\usepackage{tikz}
\usepackage{tikz-3dplot}
\usepackage{pgfplots}
\pgfplotsset{compat=1.18}

\usepackage{tikz}
\usepackage{tikz-3dplot}
\usepackage{accents}

\makeatletter
\newcommand{\doublewidetilde}[1]{{%
		\mathpalette\double@widetilde{#1}%
}}
\newcommand{\double@widetilde}[2]{%
	\sbox\z@{$\m@th#1\widetilde{#2}$}%
	\ht\z@=.9\ht\z@
	\widetilde{\box\z@}%
}
\makeatother

\newcommand{\cmark}{\textcolor{green!60!black}{\scalebox{2}{\ding{51}}}}
\newcommand{\xmark}{\textcolor{red!70!black}{\scalebox{2}{\ding{55}}}}

\usepackage{etoolbox,lipsum}

\newcommand{\blue}[1]{\textcolor{black}{#1}}
\definecolor{darkgreen}{RGB}{0,150,0}

\begin{document} 
	\title{Chirped Floquet linear drives activate forbidden charge-to-spin conversions\\ in Rashba two-dimensional electron gases}
	\author{Mohsen Yarmohammadi}
	\email{mohsen.yarmohammadi@georgetown.edu}
	\address{Department of Physics, Georgetown University, Washington DC 20057, USA}
	\date{\today}
	\begin{abstract}
		In Rashba two-dimensional electron gases~(2DEGs), charge-to-spin conversion via the Edelstein effect is conventionally limited to the transverse plane. Accessing longitudinal or out-of-plane pathways typically requires static magnetic fields or interface engineering, which cause stray fields and lack tunability. While dynamic circular or elliptical Floquet drives can also unlock these forbidden pathways, a simple Floquet linear drive cannot. Here, we propose an alternative approach: a \textit{chirped} Floquet linear drive. The chirp induces in-plane Floquet-Zeeman fields and an odd-parity momentum drift, which simultaneously break rotational and time-reversal symmetries. This mechanism activates the forbidden Edelstein charge-to-spin conversions in Rashba 2DEGs. Experimentally accessible via programmable spatial light modulators or optical delay lines, this tunable chirped linear drive offers a broadly applicable route to spin-orbit torque switching and high-efficiency spintronics.
	\end{abstract}
	
	\maketitle
	{\allowdisplaybreaks
		
		\blue{\textit{Introduction}}---The coupling between charge and spin degrees of freedom in noncentrosymmetric systems is central to modern spintronics~\cite{Manchon_2015_NatMat, Soumyanarayanan_2016_Nature,Zutic_2004_RMP,Varotto_2021_NatElectron,Vaz_2024_ACSAMI}. In two-dimensional electron gases~(2DEGs), structural inversion asymmetry induces a Rashba spin-orbit coupling~(RSOC)~\cite{Bychkov_1984_JETP,Nitta_1997_PRL,Bordoloi_2024_JAP} that links the electron spin orientation to its momentum. Under an applied DC electric field, the resulting asymmetric shift of the Fermi contour generates a net non-equilibrium spin accumulation. This phenomenon, known as the spin Edelstein effect~\cite{Edelstein_1990_SSC, Ganichev_2002_Nature,Johansson_2024}, provides an established pathway for all-electrical spin generation. 
		
		In a pristine Rashba 2DEG, continuous rotational symmetry ($C_{\infty v}$) restricts the spin Edelstein accumulation to the transverse plane~\cite{Zelezny_2017_PRB,Johansson_2021_PRB,Johansson_2024,PhysRevResearch.3.013275}, forcing the induced spin to remain orthogonal to the applied DC electric field~[Fig.~\ref{f1_new}(left)]. Because all longitudinal and out-of-plane Edelstein components vanish, conventional Rashba 2DEGs cannot provide the collinear or out-of-plane spin injection required for advanced spintronic architectures~\cite{MacNeill_2017_NatPhys,Baek_2018_NatMater,Liu_2021_NatRevPhys,Wu_2022_MatFut}. Overcoming these geometric limits typically relies on static magnetic fields or low-symmetry interfaces~\cite{Liu_2012_Science,Safeer_2019_NanoLett,Basov2017,doi:10.1142/S0217979206035680}, which precludes ultrafast, non-thermal control. While Floquet engineering offers a dynamic alternative to tailor quantum materials~\cite{Wang_2013_Science,Oka_2019_AnnuRev, Rudner_2020_NatRevPhys,delaTorre_2021_RMP,k3xb-8pts}, standard approaches face a critical bottleneck in Rashba 2DEGs: circularly or elliptically polarized light lifts structural symmetry~\cite{Lindner_2011_NatPhys,McIver_2020_NatPhys,Kitamura_2017_PRB,GomezLeon_2013_PRL,Yudin_2016_PRB,Bhattacharya_2021_PRB}, but standard linear driving cannot provide the symmetry breaking required to activate the forbidden Edelstein components~[Fig.~\ref{f1_new}(left)].
		
		In this Letter, we show that although a simple linearly polarized Floquet drive cannot activate these forbidden Edelstein components, a \textit{chirped} linear drive can~[Fig.~\ref{f1_new}(right)]. Using a high-frequency Floquet expansion, we find that this tailored optical protocol induces static, chirp-dependent in-plane Floquet-Zeeman fields and odd-parity scalar momentum drifts. These emergent fields and drifts break the intrinsic two-fold rotational ($C_{2z}$) and time-reversal ($\mathcal{T}$) symmetries of the pristine Rashba 2DEG. This chirp-induced dynamic symmetry lifting activates the forbidden diagonal and out-of-plane components of the spin Edelstein susceptibility tensor. The resulting transport regime is highly anisotropic, allowing the magnitude and three-dimensional orientation of the DC current-induced spin accumulation to be deterministically controlled purely via the optical chirp.\begin{figure}[t]
			\centering
			\includegraphics[width=0.85\linewidth]{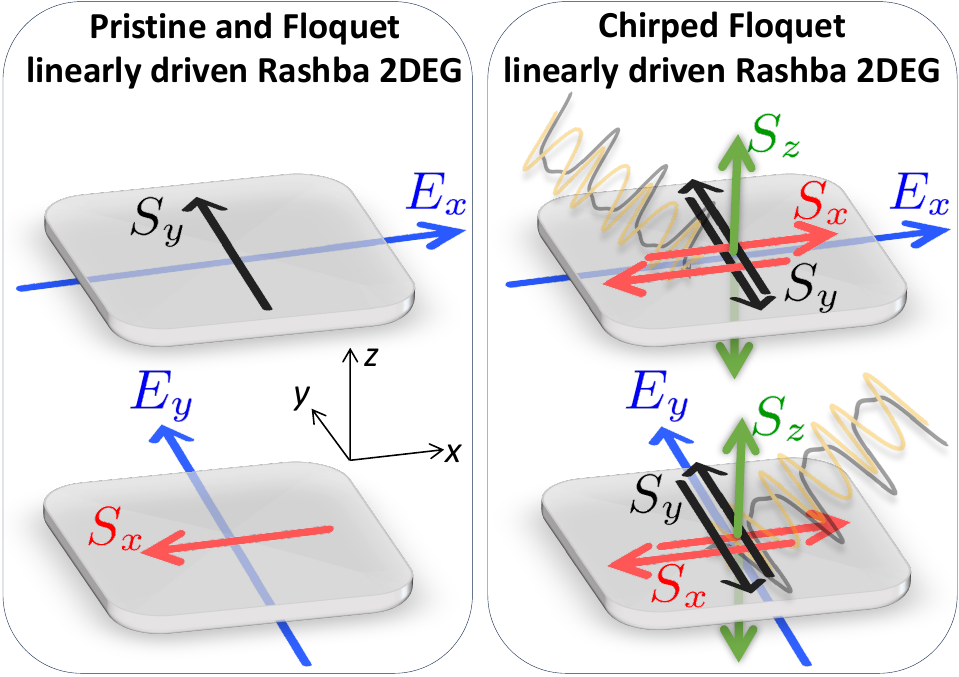}
			\caption{\textbf{Schematic of pristine and (chirped) Floquet linearly driven Rashba 2DEGs.} In the pristine Rashba 2DEG or under linealry polarized Floquet light (left), continuous symmetries restrict the DC electric field ($E_x,E_y$)-induced spin Edelstein polarization $S_{x,y}$ to be strictly orthogonal to the applied field. A chirped, linear drive (right), however, breaks the time-reversal and rotational symmetries of the system, which, in turn, lifts the native transport constraints, activating previously forbidden longitudinal and out-of-plane spin accumulations.}
			\label{f1_new}
		\end{figure}
		
		\blue{\textit{Pristine Rashba 2DEG}}---We consider a clean 2DEG in the $xy$ plane~[Fig.~\ref{f1_new}], described by the Rashba Hamiltonian $H_0(\mathbf{k}) = \epsilon k^2 \sigma_0 + \alpha(k_x\sigma_y - k_y\sigma_x)$, where $\epsilon = \hbar^2/(2m_{\rm e})$ is the kinetic energy coefficient, $\alpha$ is the RSOC strength, and $\sigma_i$ are the Pauli matrices. This interaction yields isotropic spin-split bands $\varepsilon_{\pm}(\mathbf{k}) = \epsilon k^2 \pm \alpha k$ with in-plane spin-momentum locking~[Fig.~\ref{f1}(a)]. Although the Rashba coupling breaks spatial inversion symmetry ($\mathcal{P}$), the pristine system lacks in-plane and out-of-plane $\textbf{k}$-independent mass terms, preserving both two-fold rotational ($C_{2z}$) and time-reversal ($\mathcal{T}$) symmetries. This $C_{2z}$ symmetry restricts the conventional spin Edelstein polarization strictly to the transverse plane~\cite{PhysRevResearch.3.013275}. The detailed symmetry analysis of the pristine Rashba 2DEG is provided in Sec.~S1 of the Supplemental Material~(SM)~\cite{SM}.\begin{figure}[t]
			\centering
			\includegraphics[width=1\linewidth]{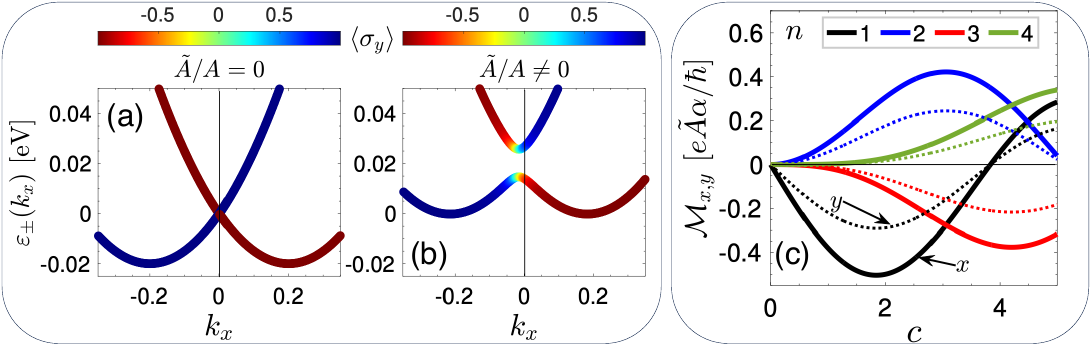}
			\caption{\textbf{Pristine and chirped Floquet quasi-energy bands, spin textures, and induced Floquet-Zeeman fields.} (a, b) Pristine and chirped Floquet quasi-energy bands with transverse spin projection $\langle \sigma_y \rangle$ (color scale) for the (a) pristine ($\tilde{A}=0$) and (b) driven ($\tilde{A}/A = 0.3$) regimes, where the drive breaks time-reversal and rotational symmetries to open an asymmetric gap at finite momentum~($k_x \neq 0$). (c) In-plane chirped Floquet-Zeeman fields $\mathcal{M}_{x,y}$ (in units of $e \tilde{A} \alpha/\hbar$, where $\alpha$ is the Rashba coupling strength) as a function of the phase chirp $c$, illustrating control over both field magnitude and sign.}
			\label{f1}
		\end{figure}
		
		\blue{\textit{Rashba 2DEG under chirped Floquet linear driving}}---To modify the transport symmetries, we employ a phase-chirped bichromatic linear drive. While a monochromatic chirped drive is also viable, the bichromatic protocol provides harmonic selectivity that enables non-linear control. We define the vector potential as $\mathbf{A}(t) = \big(A\cos(\omega t)+\tilde{A}\cos[n\omega t+c \sin(\omega t)]\cos\beta, \tilde{A}\cos[n\omega t+c \sin(\omega t)]\sin\beta\big)$, which superimposes a fundamental mode ($A$, $\omega$) with a phase-modulated $n$-th harmonic ($\tilde{A}$, phase chirp $c$, tilt angle $\beta$). Setting $A \to 0$, $\tilde{A} \to A$, and $n \to 1$ recovers the monochromatic limit. Although an amplitude-chirped drive can also introduce temporal asymmetries to break the geometric symmetries, phase-chirped protocols maintain a constant optical intensity envelope. Experimentally, this intensity stability mitigates laser-induced heating and avoids material damage thresholds.
		
		To find the steady state, we use Peierls substitution $\mathbf{k}\rightarrow \mathbf{k}+e\mathbf{A}(t)/\hbar$~\cite{doi:10.7566/JPSJ.94.111007,annurev-conmatphys-031218-013423,PhysRevResearch.4.033213,Rudner2020,Giovannini_2020}, apply a high-frequency~($\hbar \omega \approx$ 1--3 eV) Floquet expansion~\cite{Bukov04032015,RevModPhys.89.011004,PhysRevB.79.081406,annurev-conmatphys-031218-013423,Rudner2020,PhysRevB.82.235114,PhysRevX.3.031005,GRIFONI1998229,PLATERO20041}, and decompose the Hamiltonian into equilibrium, linear, and quadratic sectors: $H(\mathbf{k},t) = H_0(\mathbf{k}) + H_1(\mathbf{k},t) + H_2(\mathbf{k},t)$~(see Sec.~S2 of the SM~\cite{SM} for the derivation). The linear correction $H_1 \propto A_i(t)$ modifies the spin-momentum locking, while the quadratic Stark shift $H_2 \propto A^2_i(t)$ couples to the identity matrix $\sigma_0$. The first-order dynamical correction is governed by the commutator $[\mathcal{H}_{-m}^{\rm F}, \mathcal{H}_{m}^{\rm F}]$. Because the chirped linear protocol yields real Fourier coefficients ($A_{i,-m} = A_{i,m}$), the cross terms in the Pauli basis cancel ($\propto\left( [\sigma_x, \sigma_y] + [\sigma_y, \sigma_x]\right) = 0$). The $1/\omega$ dynamical correction therefore vanishes, leaving the effective Hamiltonian determined by the zeroth-order time average: $H_{\rm eff}(\mathbf{k}) = H_0(\mathbf{k}) + \langle H_1(\mathbf{k},t) \rangle + \langle H_2(\mathbf{k},t) \rangle$. 
		
		Expanding the nonlinear phase modulation via the Jacobi-Anger identity, the zeroth-order vector potential components are $A_{x,0} = \tilde{A}\cos\beta (-1)^n J_n(c)$ and $A_{y,0} = \tilde{A}\sin\beta (-1)^n J_n(c)$, where $J_n(c)$ is the Bessel function of the first kind of order $n$. Substituting these terms into the time averages results in the effective Hamiltonian $H_{\rm eff}(\mathbf{k}) = h_0(\mathbf{k})\sigma_0 + h_x(\mathbf{k})\sigma_x + h_y(\mathbf{k})\sigma_y$. The scalar sector,\begin{align}
			h_0(\mathbf{k}) = {} &\epsilon k^2 + \frac{2e\epsilon\tilde{A}}{\hbar}(-1)^n J_n(c) (k_x \cos\beta + k_y \sin\beta) \notag \\ {} &+ \frac{e^2\epsilon}{2\hbar^2} (A^2+\tilde{A}^2) + \frac{e^2\epsilon \tilde{A}^2}{2\hbar^2}  J_{2n}(2c) \notag \\ {} &- \frac{2n e^2\epsilon A \tilde{A} \cos\beta}{c\hbar^2} (-1)^n J_n(c)\,,
		\end{align}incorporates the kinetic energy and chirp-induced terms; an odd-parity momentum drift (the second term), and a global Stark shift. The Pauli sector contains the intrinsic spin-momentum locking modified by momentum-independent Floquet-Zeeman fields,\begin{subequations}\label{eq_2}
			\begin{align}
				h_x(\mathbf{k}) &= -\alpha k_y  -\frac{e\alpha \tilde{A}}{\hbar} (-1)^n J_n(c) \sin\beta\,,\\
				h_y(\mathbf{k}) &= \alpha k_x + \frac{e\alpha \tilde{A}}{\hbar} (-1)^n J_n(c) \cos\beta\,.
			\end{align}
		\end{subequations}When $\tilde{A} = 0$ and $c = 0$, all symmetry-breaking terms vanish. Since $J_n(0) = 0$ for $n > 0$, the induced fields, $(\mathcal{M}_x,\mathcal{M}_y) = \frac{e\alpha \tilde{A}}{\hbar} (-1)^n J_n(c) (-\sin\beta,\cos \beta)$ disappear, restoring the unperturbed spin-momentum locking. The scalar sector reduces to $h_0(\mathbf{k}) = \epsilon k^2 + e^2\epsilon A^2 / (2\hbar^2)$, where the second term acts purely as a global shift to the chemical potential. Because this rigid shift preserves both spatial inversion and time-reversal symmetries, the pristine transport constraints of the Rashba 2DEG are fully recovered. Hence, the chirp is a necessary condition for breaking these symmetries with linearly polarized light.
		
		The physical consequences of these terms are apparent in the reconstructed Floquet quasi-energy bands~[Fig.~\ref{f1}(b)]. The generation of these static fields $(\mathcal{M}_x,\mathcal{M}_y)$, alongside the odd-parity scalar momentum drift, breaks the intrinsic time-reversal ($\mathcal{T}$) and two-fold rotational ($C_{2z}$) symmetries of the pristine 2DEG (summarized in Tab.~\ref{tab1} and detailed in Sec.~S3 of the SM~\cite{SM}). Because the $\mathcal{M}_{x,y}$ fields are momentum-independent, they do not reverse sign under spatial inversion, lifting the $C_{2z}$ protection without requiring circularly or elliptically polarized light. Unlike the pristine gapless Rashba crossing, the emergent $\mathcal{M}_{x,y}$ fields and both constant and $\textbf{k}$-dependent drifts open an asymmetric energy gap at finite momentum ($k \neq 0$) and cant the unperturbed transverse spin texture. As shown in Fig.~\ref{f1}(c), tuning the optical phase chirp $c$ across the Bessel roots allows these symmetry-breaking fields to be scaled, maximized, or reversed, enabling all-optical control over the anisotropic spin Edelstein response.\begin{table}[t]
			\centering
			\resizebox{1\linewidth}{!}{
				\begin{tabular}{c|c|c}
					\hline\hline
					\makecell{Symmetry} & \makecell{Pristine and Floquet linearly\\ driven Rashba 2DEG} & \makecell{Chirped Floquet linearly \\ driven Rashba 2DEG} \\
					\hline
					Time-reversal $\mathcal{T}$ & Preserved & Broken \\\hline
					Spatial inversion $\mathcal{P}$ & Broken  & Broken \\\hline
					Two-fold rotation $C_{2z}$ & Preserved & Broken \\
					\hline\hline
			\end{tabular}}
			\caption{\textbf{Fundamental symmetry transformations.} The explicit breaking of $\mathcal{T}$ and $C_{2z}$ under the phase-chirped drive removes the geometric constraints of the pristine Rashba 2DEG, activating the anisotropic spin Edelstein response.}\label{tab1}
		\end{table}
		
		\blue{\textit{Non-equilibrium spin Edelstein polarization}}---The current-induced spin accumulation $S_i = \sum_{j} \chi_{ij} E_j$ under a DC electric field $\mathbf{E}$ is determined by evaluating the spin Edelstein susceptibility tensor $\chi_{ij}$ via the Kubo formula. Within the relaxation-time approximation, $\chi_{ij}$ splits into intraband (Fermi-surface) and interband (Fermi-sea) contributions, given by\begin{align}
			\chi_{ij} = {} &\frac{ie}{m_{\rm e}} \int \frac{d^2k}{(2\pi)^2} \Bigg[ \sum_\nu \frac{\partial f_{\nu\mathbf{k}}}{\partial \varepsilon_{\nu\mathbf{k}}} \frac{S^i_{\nu\nu,\mathbf{k}} p^j_{\nu\nu,\mathbf{k}}}{i\tau_{\rm intra}^{-1}} \notag \\
			& - \sum_{\nu\neq \zeta} \frac{f_{\nu\mathbf{k}} - f_{\zeta\mathbf{k}}}{\varepsilon_{\nu\mathbf{k}} - \varepsilon_{\zeta\mathbf{k}}} \frac{S^i_{\zeta\nu,\mathbf{k}} p^j_{\nu \zeta,\mathbf{k}}}{\varepsilon_{\zeta\mathbf{k}} - \varepsilon_{\nu\mathbf{k}} + i \tau_{\rm inter}^{-1}} \Bigg]\,,
		\end{align}where $\varepsilon_{\nu\mathbf{k}}$ and $| u_{\nu\mathbf{k}} \rangle$ are the Floquet quasi-energies and eigenstates, respectively. The spin and momentum matrix elements are defined as $S^i_{\zeta\nu,\mathbf{k}} = \langle u_{\zeta\mathbf{k}} | \sigma_i | u_{\nu\mathbf{k}} \rangle$ and $p^j_{\nu\zeta,\mathbf{k}} = m_{\rm e}\langle u_{\nu\mathbf{k}} | v_j | u_{\zeta\mathbf{k}} \rangle$. The velocity operator $v_j = \hbar^{-1}\partial H_{\rm eff}(\mathbf{k})/\partial k_j$ accounts for the drive-induced momentum drift and modified SOC. At zero temperature ($T \to 0$), the energy derivative of the Fermi-Dirac distribution simplifies the intraband term to a Fermi-surface line integral.
		
		\blue{\textit{Results and discussion}}---We evaluate the complete spin Edelstein susceptibility tensor of the driven Rashba 2DEG numerically over a dense $1000 \times 1000$ $\mathbf{k}$-space grid. Working in natural units ($e = m_{\rm e} = \hbar = k_B = 1$), the integration includes the $(2\pi)^{-2}$ density of states factor. To isolate intrinsic geometric band effects from thermal smearing, the system is evaluated in the low-temperature limit ($k_B T = 1$~meV) with a Gaussian broadening of $\eta = 0.01$~eV and relaxation rates of $\hbar/\tau = 0.1$~eV. We present the results in terms of the dimensionless susceptibilities $\widetilde{\chi}_{ij}$, where the transverse components $\widehat{\chi}_{ij}$ are normalized to their unperturbed Rashba baseline ($\widehat{\chi}_{ij} = \widetilde{\chi}_{ij} / \widetilde{\chi}_{xy}|_{\rm Eq.}$).\begin{table}[t]
			\centering
			\resizebox{1\linewidth}{!}{\begin{tabular}{c|c|c}
					\hline\hline
					\makecell{Edelstein\\ component} & \makecell{Pristine and Floquet\\ linearly driven Rashba 2DEG} & \makecell{Chirped Floquet \\ linearly driven Rashba 2DEG} \\
					\hline
					$\chi_{xx}$ 
					& \xmark
					& \cmark
					\\\hline
					
					$\chi_{yy}$
					& \xmark
					& \cmark
					\\\hline
					
					$\chi_{xy}$
					& \cmark
					& \cmark
					\\\hline
					
					$\chi_{yx}$
					& \cmark ($ = -\chi_{xy}$)
					&  \cmark ($ \neq -\chi_{xy}$)
					\\\hline
					
					$\chi_{zx}$
					& \xmark 
					& \cmark
					\\\hline
					
					$\chi_{zy}$
					& \xmark 
					& \cmark 
					\\
					\hline\hline
			\end{tabular}}
			\caption{\textbf{Symmetry-allowed spin Edelstein susceptibilities with and without chirped Floquet linear drivings.} In the static Rashba 2DEG or under non-chirped linear drive, continuous symmetries restrict the spin accumulation to the transverse plane, forcing an antisymmetric tensor ($\chi_{yx} = -\chi_{xy}$) and forbidding all diagonal ($xx, yy$) and out-of-plane ($zx, zy$) responses. Under the phase-chirped linear drive, the breaking of rotational and time-reversal symmetries activates all previously forbidden tensor components and lifts the transverse antisymmetry.}	\label{tab2}
		\end{table} 
		
		Evaluating the Kubo formula over the unperturbed bands yields an antisymmetric tensor ($\chi_{yx} = -\chi_{xy} \neq 0$), while all diagonal and out-of-plane components vanish ($\chi_{xx} = \chi_{yy} = \chi_{zx} = \chi_{zy} = 0$)~\cite{PhysRevResearch.3.013275}. The phase-chirped linear drive modifies this behavior by distorting the Fermi contours. The emergent in-plane fields $\mathcal{M}_{x,y}$ break the $C_{2z}$ rotational symmetry, canting the spin expectation values $S^i_{\nu\nu,\mathbf{k}}$ away from orthogonality with the velocity. Displacing this asymmetric Fermi surface under a DC field yields finite, tunable diagonal and out-of-plane components whose magnitudes depend directly on the phase chirp $c$ and polarization tilt angle $\beta$. These dynamically activated tensor components are summarized in Tab.~\ref{tab2}
		
		Decomposing the susceptibility reveals distinct microscopic transport mechanisms for the different tensor components. The induced in-plane spin accumulations are dissipative, governed by the intraband (Fermi-surface) displacement of the Floquet-modified contours under the DC field~[Fig.~\ref{f2}(a,b)]. Conversely, the out-of-plane spin polarization emerges from the interband (Fermi-sea) susceptibility~[Fig.~\ref{f2}(c)]. Because the 2D intraband expectation values confine the spins to the plane ($S^z_{\nu\nu,\mathbf{k}} = 0$), out-of-plane polarization requires non-vanishing interband matrix elements ($S^z_{\nu\nu',\mathbf{k}} \neq 0$), which are enabled here by the lifting of $\mathcal{T}$ and rotational symmetries. The applied DC field, thus, drives an interband mixing that generates out-of-plane charge-to-spin conversion via a quantum coherence pathway.
		
		Figures~\ref{f2}(d--i) show the activation of the forbidden charge-to-spin conversion pathways under the chirped drive. In the limit $\tilde{A}/A \to 0$, the pristine system is recovered: the diagonal~[Figs.~\ref{f2}(d,e)] and out-of-plane~[Figs.~\ref{f2}(h,i)] components vanish, while the transverse responses~[Figs.~\ref{f2}(f,g)] return to the unperturbed baseline. Increasing $\tilde{A}$ breaks the $C_{2z}$ and $\mathcal{T}$ symmetries, inducing the previously forbidden $\widetilde{\chi}_{xx}$, $\widetilde{\chi}_{yy}$, $\widetilde{\chi}_{zx}$, and $\widetilde{\chi}_{zy}$ susceptibilities. Figures~\ref{f2}(d,f,h) illustrates the non-monotonic scaling of this symmetry breaking. At a fixed harmonic order ($n=2$), the emergent fields are proportional to the Bessel function $J_2(c)$. Because $|J_2(c)|$ peaks near $c \approx 3.05$, the transport response is maximized at $c=3$ before decreasing at $c=4$. On the other hand, Figs.~\ref{f2}(e,g,i) map the harmonic coupling efficiency at a fixed chirp $c=2$. The transport signatures follow the Bessel function hierarchy $J_1(2) > J_2(2) > J_3(2) \gg J_4(2)$, confirming that lower-harmonic mixing provides stronger modifications to the Rashba spin texture.\begin{figure}[t]
			\centering
			\includegraphics[width=0.85\linewidth]{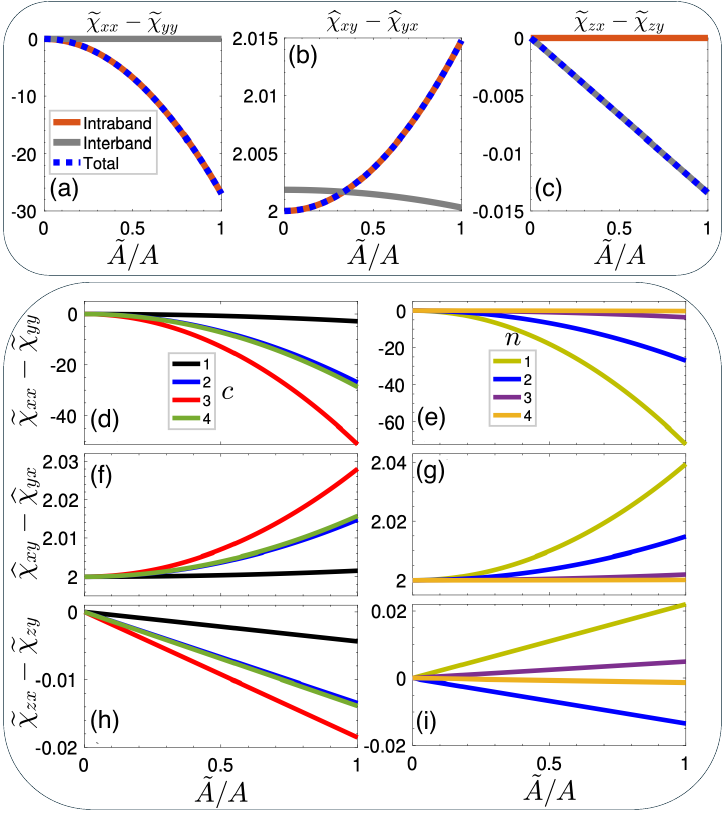}
			\caption{\textbf{Chirped Floquet-induced charge-to-spin conversion in Rashba 2DEGs.} (a–c) Decomposition of the dynamically induced (a) longitudinal, (b) transverse, and (c) out-of-plane susceptibilities into intraband, interband, and total contributions as a function of the drive amplitude ratio $\tilde{A}/A$ ($c=2, n=2$). The in-plane accumulations arise from intraband Fermi-surface displacements, whereas out-of-plane spin generation originates from interband quantum coherence. (d–i) Evolution of the induced total tensor components versus $\tilde{A}/A$ for (d, f, h) varying phase chirp $c$ at fixed $n=2$, and (e, g, i) varying harmonic orders $n$ at fixed $c=2$. Parameters: $\alpha = 0.1$ eV\,\AA, $eA/\hbar k_F = 0.3$, and $\beta = \pi/3$.}
			\label{f2}
		\end{figure}
		
		Figure~\ref{f2}(i) shows a sign reversal in the out-of-plane susceptibilities depending on the harmonic order; while $n=1$ yields a positive slope, $n=2$ inverts the response due to the $(-1)^n$ parity factor governing the emergent Floquet-Zeeman fields in Eq.~\eqref{eq_2}. This tunability establishes an anisotropic transport regime where spin generation depends on the relative angle between the applied field $\mathbf{E}$ and the induced field $\boldsymbol{\mathcal{M}}$. Rather than being constrained by the native crystal geometry, the non-equilibrium spin polarization vector can be continuously rotated across transverse, longitudinal, and out-of-plane directions by tuning the optical phase chirp $c$.\begin{figure}[t]
			\centering
			\includegraphics[width=0.85\linewidth]{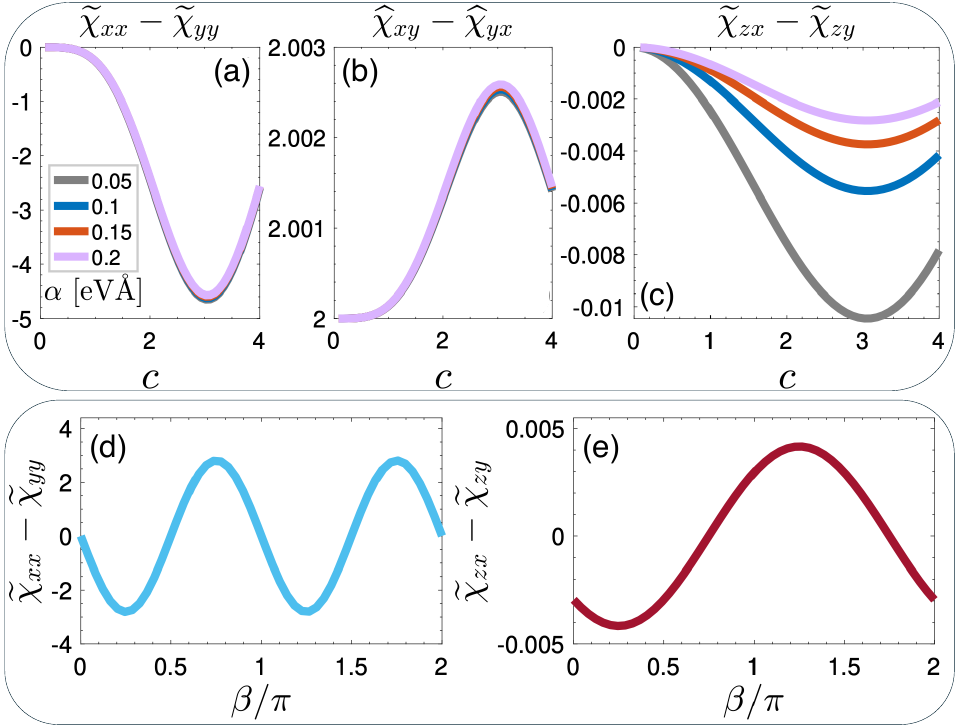}
			\caption{\textbf{Chirped Floquet-induced Edelstein susceptibilities as a function of Rashba coupling strength and polarization angle.} (a–c) Evolution of the dynamically induced (a) longitudinal, (b) transverse, and (c) out-of-plane spin Edelstein susceptibilities versus the phase-chirp parameter $c$ for various RSOC strengths $\alpha$ ($\tilde{A}/A = 1, n = 2, \beta = \pi/3$). (d, e) Dependence of the induced (d) in-plane diagonal and (e) out-of-plane susceptibilities on the polarization tilt angle $\beta$ at $c=2$ and $n=2$, showing $\pi$-periodic and $2\pi$-periodic behaviors, respectively. Parameters: $\tilde{A}/A = 0.3$ and $eA/\hbar k_F = 0.3$.}
			\label{f3}
		\end{figure}
		
		For a representative longitudinal dimensionless susceptibility of $\chi = 10$ under an applied DC electric field of $E = 1$~V/nm, the system generates a local spin accumulation density of $10\,\mu_{\rm B}/\text{nm}^2$. Integrating this density over a standard 2D lattice unit cell area ($a^2 \approx 0.01\,\text{nm}^2$, corresponding to the model's momentum cutoff $\pi/a$), the induced polarization equals approximately $0.1\,\mu_{\rm B}$ per unit cell. 
		
		Figure~\ref{f3}(a--c) presents the relationship between the chirp parameter $c$ and the intrinsic RSOC strength $\alpha$. Across all configurations, the susceptibilities exhibit a non-monotonic profile governed by the $|J_2(c)|$ Bessel scaling that peaks at $c \approx 3$. The in-plane transport—encompassing both the longitudinal~[Fig.~\ref{f3}(a)] and transverse~[Fig.~\ref{f3}(b)] components—is independent of $\alpha$, with curves for different Rashba strengths overlapping. Physically, this stems from a compensation effect: a larger $\alpha$ increases the spin splitting but also increases the rigidity of the spin texture against Fermi contour displacements. Within linear response, this rigidity cancels the changes in the density of states, leaving the in-plane response determined by the optical drive. However, the out-of-plane susceptibilities~[Fig.~\ref{f3}(c)] decrease as $\alpha$ increases. Because the unperturbed Rashba interaction acts as an in-plane field that locks the spins to the plane, a stronger $\alpha$ suppresses the out-of-plane canting induced by the drive. Thus, host materials with weaker SOC are preferred for optimizing out-of-plane spin generation.
		
		Figure~\ref{f3}(d,e) shows the dynamically activated longitudinal and out-of-plane responses as a function of the polarization tilt angle $\beta$. The longitudinal components exhibit a $\pi$-periodicity [Fig.~\ref{f3}(d)] arising from the lowest-order mixing of the orthogonal field components ($\mathcal{M}_x \propto \sin\beta$, $\mathcal{M}_y \propto \cos\beta$). In contrast, the out-of-plane responses display a $2\pi$-periodicity [Fig.~\ref{f3}(e)], as the out-of-plane canting depends on the orientation of the induced Floquet-Zeeman field relative to the momentum drift. This angular dependence provides a means for three-dimensional spin vector control: rotating the incident polarization plane tunes the in-plane components while independently reversing the sign of the out-of-plane spin accumulation.
		
		\blue{\textit{Experimental feasibility}}---An experimental realization of this protocol involves a phase-locked, two-color near-infrared laser coupled to a high-mobility Rashba 2DEG, such as gate-tunable $\text{LaAlO}_3/\text{SrTiO}_3$ interfaces~\cite{PhysRevLett.104.126803} or $\text{GaAs/AlGaAs}$ heterostructures~\cite{PhysRevLett.90.076807,PhysRevLett.92.256601,Meier2007,Koralek2009}. High-mobility samples are needed to preserve these modified spin textures against scattering. While the activated in-plane spin generation is independent of $\alpha$, maximizing the out-of-plane spin component favors a weak Rashba platform ($\alpha \approx 0.05\ \text{eV\,\AA}$) to reduce the intrinsic planar spin-locking. To suppress dissipative interband absorption, the drive can utilize a fundamental wavelength of $10\ \mu\text{m}$ ($\hbar \omega \approx 124\text{ meV}$) synchronized with its second harmonic $20\ \mu\text{m}$ ($n=2$). Reaching the driving strength $eA/\hbar k_F \approx 0.3$ corresponds to a peak in-plane electric field of $\approx 36\text{ kV/cm}$ (intensity $\sim 1.7\text{ MW/cm}^2$), which can be delivered via low-repetition-rate picosecond pulses to prevent thermal damage. The phase-chirp parameter (e.g., $c \approx 3.05$) can be controlled using a sub-wavelength optical delay line. The resulting longitudinal spin components can be detected via non-local spin valves or magneto-optical Kerr effect (MOKE) microscopy~\cite{doi:10.1126/science.1105514,Lou2007,Sanchez2013}. Finally, toggling the harmonic order $n$ optically flips the sign of the out-of-plane polarization via the $(-1)^n$ scaling, providing an experimental signature of the underlying Floquet mechanism.
		
		\blue{\textit{Conclusion}}---In summary, while circular or elliptical drives generate multi-directional Edelstein charge-to-spin conversions in Rashba 2DEGs, standard linear Floquet drives restrict these conversions strictly to the axis orthogonal to the applied DC electric field. Here, we propose an alternative method to induce the otherwise forbidden pathways using a chirped linear Floquet drive. Within a high-frequency Floquet framework, this protocol generates stationary in-plane Floquet-Zeeman fields and an odd-parity momentum drift. The resulting dynamic symmetry lifting breaks time-reversal and rotational invariance, thereby activating the previously forbidden Edelstein channels. By circumventing the need for static magnetic fields or tailored interfaces, this material-agnostic chirp-based Floquet approach eliminates magnetic cross-talk and provides a versatile platform for field-free spin-orbit torque switching in high-density spintronics.
		
		Future directions include extending this protocol to the orbital Edelstein effect and non-equilibrium Hall responses~\cite{yarmohammadi2026floquetinducedanisotropicmagnetoresistanceanomalous,yarmohammadi2026giantperpendicularedelsteinpolarization}. Furthermore, applying this chirped drive to systems with higher-order SOC and finite spin splittings~\cite{doi:10.1126/sciadv.aaz8809, hayami2019momentum, PhysRevX.12.031042, PhysRevX.12.040501, bai2024altermagnetism, liu2025different, krempasky2024altermagnetic, lee2024broken, osumi2024observation, yarmohammadi2026floquetinducedanisotropicmagnetoresistanceanomalous, yarmohammadi2026efficienttwocolorfloquetcontrol,yarmohammadi2026giantperpendicularedelsteinpolarization} could reveal alternative phenomena for all-optical spintronic and orbitronic logic architectures.
		
		\blue{\textit{Acknowledgments}}---M.\,Y. gratefully thanks useful discussions with Peter M. Oppeneer and Libor \v{S}mejkal and acknowledges the hospitality of Uppsala University during his visit, where this work was performed. M.\,Y. was supported by the Department of Energy, Office of Basic Energy Sciences, Division of Materials Sciences and Engineering under Contract No. DE-FG02-08ER46542 for the formal developments, the analytical/numerical work, and writing of the manuscript. 
		
		
	}
	
	\bibliography{bib.bib}
	
\onecolumngrid
\clearpage

{\allowdisplaybreaks
	
	\begin{center}
		\textbf{\large \vskip0mm Supplemental Materials for ``Chirped Floquet linear drives activate forbidden charge-to-spin conversions in Rashba two-dimensional electron gases'}
		\vskip3.5mm
		Mohsen Yarmohammadi$^1$\vskip1mm
		\small $^1$\textit{Department of Physics, Georgetown University, Washington DC 20057, USA}\\
		(Dated: \today)
	\end{center}
	\setcounter{equation}{0}

	\makeatletter

	\setcounter{equation}{0}
	\renewcommand{\theequation}{S\arabic{equation}}
	\setcounter{figure}{0}
	\renewcommand{\thefigure}{S\arabic{figure}}
	\setcounter{section}{0}
	\renewcommand{\thesection}{S\arabic{section}}
	\setcounter{table}{0}
	\renewcommand{\thetable}{S\arabic{table}}
	
	\section{S1. Symmetry Analysis of the Unperturbed Hamiltonian}
	To understand the physical impact of the chirped Floquet linearly polarized drive, we must first establish the symmetry properties of the unperturbed Rashba 2DEG Hamiltonian $H_0(\mathbf{k}) = \epsilon(k_x^2+k_y^2)\sigma_0 + \alpha(k_x\sigma_y-k_y\sigma_x)$.
	
	\subsection*{S1.1. Time-reversal symmetry ($\mathcal{T}$)}
	The time-reversal operator for a spin-$\frac{1}{2}$ system is defined as $\mathcal{T} = -i\sigma_y \mathcal{K}$, where $\mathcal{K}$ is complex conjugation. Under time-reversal, momentum reverses ($\mathbf{k} \rightarrow -\mathbf{k}$) and the Pauli matrices transform as $\mathcal{T} \boldsymbol{\sigma} \mathcal{T}^{-1} = -\boldsymbol{\sigma}$.
	Applying this to the unperturbed Hamiltonian gives 
	\begin{align}
		\mathcal{T} H_0(\mathbf{k}) \mathcal{T}^{-1} = \epsilon(k_x^2+k_y^2)\sigma_0 + \alpha \big(k_x(-\sigma_y) - k_y(-\sigma_x)\big)= \epsilon(k_x^2+k_y^2)\sigma_0 + \alpha \big(-k_x\sigma_y + k_y\sigma_x\big)\,.
	\end{align}
	Evaluating the Hamiltonian at the reversed momentum yields\begin{equation}
		H_0(-\mathbf{k}) = \epsilon\big((-k_x)^2+(-k_y)^2\big)\sigma_0 + \alpha\big((-k_x)\sigma_y - (-k_y)\sigma_x\big)  \,.
	\end{equation}
	Comparing the two, $\mathcal{T} H_0(\mathbf{k}) \mathcal{T}^{-1} = H_0(-\mathbf{k})$. Therefore, time-reversal symmetry is preserved in a pristine Rashba 2DEG.
	
	\subsection*{S1.2. Spatial inversion symmetry ($\mathcal{P}$)}
	The spatial inversion operator in this 2D plane does not flip the spin (which is a pseudovector), so $\mathcal{P} = \sigma_0$. Under inversion, $\mathbf{k} \rightarrow -\mathbf{k}$, we have $\mathcal{P} H(\mathbf{k}) \mathcal{P}^{-1} = H(\mathbf{k})$. However, looking at the reversed momentum state leads to
	\begin{equation}
		H(-\mathbf{k}) = \epsilon k^2\sigma_0 - \alpha(k_x\sigma_y - k_y\sigma_x) \neq H(\mathbf{k})\,.
	\end{equation}
	The presence of the linear Rashba term explicitly dictates that spatial inversion symmetry is broken, reflecting the structural asymmetry defining the Rashba effect.

	\subsection*{S1.3. Two-fold rotation symmetry ($C_{2z}$)}
	Let us also test a $180^\circ$ rotation around the $z$-axis. This maps $(x, y) \rightarrow (-x, -y)$ and $\mathbf{k} \rightarrow -\mathbf{k}$. Spin rotates by $\pi$, given by the operator $\mathcal{U}_{C2z} = -i\sigma_z$. This flips both $\sigma_x$ and $\sigma_y$, leading to
	\begin{align}
		\mathcal{U}_{C2z} H(\mathbf{k}) \mathcal{U}_{C2z}^{-1} = \epsilon k^2\sigma_0 + \alpha\big(k_x(-\sigma_y) - k_y(-\sigma_x)\big) = \epsilon k^2\sigma_0 - \alpha\big(k_x\sigma_y - k_y\sigma_x\big)\,.
	\end{align}This matches $H(-\mathbf{k})$ perfectly. Therefore, the unperturbed Rashba 2DEG preserves $C_{2z}$ rotational symmetry (and, by extension, full continuous $C_{\infty z}$ rotational symmetry).
	
	\section{S2. Derivation of the Effective Floquet Hamiltonian}
	We begin with the unperturbed Hamiltonian of a Rashba 2DEG, given by\begin{equation}
		H_0(\mathbf{k}) = \epsilon(k_x^2+k_y^2)\sigma_0 + \alpha(k_x\sigma_y-k_y\sigma_x) \, .
	\end{equation}We aim to derive the effective Floquet Hamiltonian up to first order in the high-frequency expansion ($1/\omega$), $\mathcal{H}_{\rm eff}(\mathbf{k})  = \mathcal{H}_0^{\rm F}(\mathbf{k})  + \sum_{m\neq0} \frac{[\mathcal{H}_{-m}^{\rm F}(\mathbf{k}) ,\mathcal{H}_{+m}^{\rm F}(\mathbf{k}) ]}{m\hbar\omega}$, where the Floquet Fourier components are given by the time-average over one period $T = 2\pi/\omega$, i.e.,\begin{equation}
		\mathcal{H}_{m}^{\rm F}(\mathbf{k}) = \frac{\omega}{2\pi} \int_{0}^{2\pi/\omega} dt\; H(\mathbf{k},t) e^{im\omega t}\, .
	\end{equation}The phase-chirped vector potential $\mathbf{A}(t) = (A_x(t), A_y(t))$ driving the system is defined as\begin{subequations}
		\begin{align}
			A_x(t) &= A\cos(\omega t) + \tilde{A}\cos[n\omega t+c \sin(\omega t)] \cos\beta\,, \\
			A_y(t) &=  \tilde{A}\cos[n\omega t+c \sin(\omega t)] \sin\beta\,.
		\end{align}
	\end{subequations}To apply Floquet theory, we must decompose the vector potential into its Fourier series, $\mathbf{A}(t) = \sum_{m=-\infty}^{\infty} \mathbf{A}_m e^{-im\omega t}$. We expand the highly non-linear phase-modulated term using the Jacobi-Anger identity, $e^{ic\sin(\omega t)} = \sum_{l=-\infty}^{\infty} J_l(c) e^{il\omega t}$. Using Euler's formula $\cos(\theta) = \frac{1}{2}(e^{i\theta} + e^{-i\theta})$, the phase-chirped drive becomes\begin{align}
		\cos[n\omega t + c\sin(\omega t)] = \frac{1}{2} \left( e^{in\omega t}e^{ic\sin(\omega t)} + e^{-in\omega t}e^{-ic\sin(\omega t)} \right) = \frac{1}{2} \sum_{l=-\infty}^{\infty} J_l(c) e^{i(n+l)\omega t} + \frac{1}{2} \sum_{l=-\infty}^{\infty} J_l(-c) e^{-i(n-l)\omega t}\,.
	\end{align}Using the Bessel function property $J_l(-c) = (-1)^l J_l(c)$, this simplifies to\begin{align}
		\cos[n\omega t + c\sin(\omega t)] = \frac{1}{2}\sum_{l=-\infty}^{\infty} J_l(c) \left( e^{i(n+l)\omega t} + (-1)^l e^{-i(n-l)\omega t} \right)\,.
	\end{align}
	
	The $m=0$ DC component, $\mathbf{A}_0$, is the time-average, shown by $\langle \dots \rangle$, of the field. We extract this by finding the terms in the expansion where the exponent is zero. 
	For the first term $e^{i(n+l)\omega t}$, this requires $l = -n$.
	For the second term $e^{-i(n-l)\omega t}$, this requires $l = n$.
	Substituting these indices into the coefficients yields\begin{align}
		\langle \cos[n\omega t + c\sin(\omega t)] \rangle = \frac{1}{2} \left( J_{-n}(c) + (-1)^n J_n(c) \right)\,.
	\end{align}Using the identity $J_{-n}(c) = (-1)^n J_n(c)$, the two terms are identical, leading to $\langle \cos[n\omega t + c\sin(\omega t)] \rangle = (-1)^n J_n(c)$. Thus, the zeroth-order Fourier components are\begin{subequations}\begin{align}
			A_{x,0} &= \tilde{A}\cos\beta (-1)^n J_n(c)\,, \\
			A_{y,0} &= \tilde{A}\sin\beta (-1)^n J_n(c)\,.
	\end{align}\end{subequations}
	
	For all other modes, we match the exponents to $e^{-im\omega t}$. This allows us to gather the coefficients into a strictly real scalar variable $X_m$,\begin{equation}
		X_m = \frac{1}{2}\left[ (-1)^{n+m}J_{n+m}(c) + (-1)^{n-m}J_{n-m}(c) \right]\,.
	\end{equation}The discrete Fourier components for $m \neq 0$ are therefore\begin{subequations}\begin{align}
			A_{x,m} &= \frac{A}{2}(\delta_{m,1} + \delta_{m,-1}) + \tilde{A}X_m \cos\beta\,, \\
			A_{y,m} &= \tilde{A}X_m \sin\beta\,.
	\end{align}\end{subequations}Notice that $X_{-m} = X_m$. Because there are no imaginary units ($i$) in these expressions, all Fourier coefficients are purely real numbers. Therefore, $A_{i,-m} = A_{i,m}^* = A_{i,m}$.
	
	We next introduce the electromagnetic field into the unperturbed Hamiltonian via the minimal coupling substitution $\mathbf{k}\rightarrow\mathbf{k}+\frac{e}{\hbar}\mathbf{A}(t)$, given by\begin{equation}
		H(\mathbf{k},t) = \epsilon \left[ \left(k_x + \frac{e}{\hbar}A_x(t)\right)^2 + \left(k_y + \frac{e}{\hbar}A_y(t)\right)^2 \right]\sigma_0 + \alpha \left[ \left(k_x + \frac{e}{\hbar}A_x(t)\right)\sigma_y - \left(k_y + \frac{e}{\hbar}A_y(t)\right)\sigma_x \right] \,.
	\end{equation}Substituting expanded squares into the Hamiltonian and grouping by Pauli matrices gives $H(\mathbf{k},t) = H_0(\mathbf{k}) + H_1(\textbf{k},t) + H_2(\textbf{k},t)$, where\begin{subequations}\begin{align}
			H_1(\textbf{k},t) &= \left[ \frac{2e\epsilon}{\hbar} \big(k_x A_x(t) + k_y A_y(t)\big) \right] \sigma_0 + \left[ \frac{e\alpha}{\hbar} A_x(t) \right] \sigma_y - \left[ \frac{e\alpha}{\hbar} A_y(t) \right] \sigma_x \,,\\
			H_2(\textbf{k},t) &= \left[ \frac{e^2\epsilon}{\hbar^2} \big(A_x^2(t) + A_y^2(t)\big) \right] \sigma_0\,.
	\end{align}\end{subequations}
	The Fourier component $\mathcal{H}_m^{\rm F}(\textbf{k})$ for any non-zero mode maps directly to the linear and quadratic components derived above, i.e.,\begin{equation}
		\mathcal{H}_m^{\rm F}(\textbf{k}) = \underbrace{\left[ \frac{2e\epsilon}{\hbar}(k_x A_{x,m} + k_y A_{y,m}) + \frac{\epsilon e^2}{\hbar^2}(A^2)_m \right]}_{C_m(\textbf{k})} \sigma_0 + \underbrace{\frac{e\alpha}{\hbar} A_{x,m}}_{V_{y,m}} \sigma_y + \underbrace{\left(-\frac{e\alpha}{\hbar} A_{y,m}\right)}_{V_{x,m}} \sigma_x\,.
	\end{equation}We need to calculate the commutator $[\mathcal{H}_{-m}^{\rm F}(\textbf{k}), \mathcal{H}_{m}^{\rm F}(\textbf{k})]$.
	Note that the identity matrix $\sigma_0$ commutes with all Pauli matrices: $[\sigma_0, \sigma_x] = 0$, $[\sigma_0, \sigma_y] = 0$, and $[\sigma_0, \sigma_z] = 0$. Thus, the scalar sector $C_m(\textbf{k})$ produces zero cross-terms and drops out entirely, leading to\begin{equation}
		[\mathcal{H}_{-m}^{\rm F}, \mathcal{H}_{m}^{\rm F}] = [V_{x,-m}\sigma_x + V_{y,-m}\sigma_y, \; V_{x,m}\sigma_x + V_{y,m}\sigma_y]\,.
	\end{equation}Using the reality condition $A_{i,-m} = A_{i,m}$, we deduce that the vector coefficients are symmetric: $V_{x,-m} = V_{x,m}$ and $V_{y,-m} = V_{y,m}$.
	Substituting these into the commutator yields\begin{equation}
		[\mathcal{H}_{-m}^{\rm F}, \mathcal{H}_{m}^{\rm F}] = [V_{x,m}\sigma_x + V_{y,m}\sigma_y, \; V_{x,m}\sigma_x + V_{y,m}\sigma_y]\,.
	\end{equation}Let us expand this using the linearity of commutators $[A+B, C+D] = [A,C] + [A,D] + [B,C] + [B,D]$ as \begin{align}
		[\mathcal{H}_{-m}^{\rm F}, \mathcal{H}_{m}^{\rm F}] &= V_{x,m} V_{x,m} [\sigma_x, \sigma_x] + V_{x,m} V_{y,m} [\sigma_x, \sigma_y] + V_{y,m} V_{x,m} [\sigma_y, \sigma_x] + V_{y,m} V_{y,m} [\sigma_y, \sigma_y]\,.
	\end{align}Since any matrix commutes with itself, $[\sigma_x, \sigma_x] = 0$ and $[\sigma_y, \sigma_y] = 0$. Using the fundamental Pauli commutation relation $[\sigma_x, \sigma_y] = 2i\sigma_z$ and $[\sigma_y, \sigma_x] = -2i\sigma_z$, we obtain
	\begin{equation}
		[\mathcal{H}_{-m}^{\rm F}, \mathcal{H}_{m}^{\rm F}] = V_{x,m} V_{y,m} (2i\sigma_z) + V_{y,m} V_{x,m} (-2i\sigma_z) = 0\,.
	\end{equation}Because the commutator evaluates exactly to zero for all $m$, the entire first-order Floquet correction vanishes.
	
	Since the $1/\omega$ correction is zero, the effective Hamiltonian is dictated entirely by the time-average of the Hamiltonian over one optical cycle, $\mathcal{H}_{\rm eff}(\textbf{k}) = \mathcal{H}_0^{\rm F} = H_0(\mathbf{k}) + \langle H_1(t) \rangle + \langle H_2(t) \rangle$. We now substitute the non-zero DC components $A_{x,0}$ and $A_{y,0}$ into $H_1$ via\begin{equation}
		\langle H_1(t) \rangle = \frac{2e\epsilon}{\hbar} \Big(k_x A_{x,0} + k_y A_{y,0}\Big)\sigma_0 + \frac{e\alpha}{\hbar} \Big(A_{x,0}\sigma_y - A_{y,0}\sigma_x\Big)\,.
	\end{equation}Plugging in the exact expressions for $A_{x,0}$ and $A_{y,0}$ results in\begin{align}
		\langle H_1(t) \rangle =& \left[\frac{2e\epsilon}{\hbar} \Big( k_x \tilde{A}\cos\beta (-1)^n J_n(c) + k_y \tilde{A}\sin\beta (-1)^n J_n(c) \Big) \right]\sigma_0 \nonumber \\
		&+ \frac{e\alpha}{\hbar} \Big( \tilde{A}\cos\beta (-1)^n J_n(c) \sigma_y - \tilde{A}\sin\beta (-1)^n J_n(c) \sigma_x \Big)\,.
	\end{align}Factoring out the common terms, we achieve\begin{align}
		\langle H_1(t) \rangle =\left[\frac{2e\epsilon \tilde{A}}{\hbar} (-1)^n J_n(c) \big(k_x \cos\beta + k_y \sin\beta\big)\right]\sigma_0 + \frac{e\alpha \tilde{A}}{\hbar} (-1)^n J_n(c) \big(\cos\beta \sigma_y - \sin\beta \sigma_x\big)\,.
	\end{align}
	We then compute the time average of the squared vector potential, $\langle A_x^2(t) + A_y^2(t) \rangle$.
	First, we expand the squares,\begin{subequations}\begin{align}
			A_x^2(t) ={}& A^2\cos^2(\omega t) + 2A \tilde{A}\cos\beta \cos(\omega t)\cos[n\omega t+c \sin(\omega t)] + \tilde{A}^2\cos^2\beta \cos^2[n\omega t+c \sin(\omega t)]\,, \\
			A_y^2(t) ={}& \tilde{A}^2\sin^2\beta \cos^2[n\omega t+c \sin(\omega t)]\,.
	\end{align}\end{subequations}Summing them and using the identity $\sin^2\beta + \cos^2\beta = 1$, we find\begin{equation}
		A^2(t) = A^2\cos^2(\omega t) + \tilde{A}^2\cos^2[n\omega t+c \sin(\omega t)] + 2A\tilde{A}\cos\beta \cos(\omega t)\cos[n\omega t+c \sin(\omega t)]\,.
	\end{equation}
	
	Now we evaluate the time-average of each specific term analytically. The simple harmonic square is $\langle \cos^2(\omega t) \rangle = \frac{1}{2}$. For the phase-modulated square, we apply the double-angle trigonometric identity $\langle \cos^2[n\omega t+c \sin(\omega t)] \rangle = \frac{1}{2} + \frac{1}{2} \langle \cos[2n\omega t + 2c\sin(\omega t)] \rangle$. Using the DC component calculation, where the parameter $c \rightarrow 2c$ and $n \rightarrow 2n$, the time average of the remaining oscillating term is simply $(-1)^{2n} J_{2n}(2c)$. Since $(-1)^{2n} = 1$, we obtain
	\begin{equation}
		\langle \cos^2[n\omega t+c \sin(\omega t)] \rangle = \frac{1}{2} + \frac{1}{2} J_{2n}(2c)\,.
	\end{equation}
	For the cross-term, we use the trigonometric product-to-sum formula $\cos(A)\cos(B) = \frac{1}{2}[\cos(A+B) + \cos(A-B)]$ in $\cos(\omega t)\cos[n\omega t+c \sin(\omega t)] = \frac{1}{2} \cos[(n+1)\omega t + c\sin(\omega t)] + \frac{1}{2} \cos[(n-1)\omega t + c\sin(\omega t)]$, leading to $\langle \cos(\omega t)\cos[n\omega t+c \sin(\omega t)] \rangle = \frac{1}{2} \left[ (-1)^{n+1} J_{n+1}(c) + (-1)^{n-1} J_{n-1}(c) \right]$.	Because $(-1)^{n+1} = (-1)^{n-1} \cdot (-1)^2 = (-1)^{n-1}$, we can factor it out as $\frac{1}{2} (-1)^{n-1} \left[ J_{n+1}(c) + J_{n-1}(c) \right]$.	Finally, we apply the fundamental Bessel function recurrence relation $J_{n-1}(c) + J_{n+1}(c) = \frac{2n}{c}J_n(c)$, leading to\begin{equation}
		\langle \cos(\omega t)\cos[n\omega t+c \sin(\omega t)] \rangle = \frac{1}{2} (-1)^{n-1} \left( \frac{2n}{c} J_n(c) \right) = (-1)^{n-1} \frac{n}{c} J_n(c)\,.
	\end{equation}
	Multiplying these evaluated integrals by their respective coefficients from $A^2(t)$ gives the exact scalar shift $\langle A^2 \rangle$, \begin{equation}
		\langle A^2 \rangle = \frac{A^2}{2} + \frac{\tilde{A}^2}{2}\Big[1 + J_{2n}(2c)\Big] + A \tilde{A}\cos\beta \left[ (-1)^{n-1} \frac{2n}{c} J_n(c) \right]\,.
	\end{equation}
	
	If we compile all parts into the final effective Floquet Hamiltonian, organized by Pauli matrices, we find $\mathcal{H}_{\rm eff}(\mathbf{k}) = h_0(\mathbf{k})\sigma_0 + h_x(\mathbf{k})\sigma_x + h_y(\mathbf{k})\sigma_y$, where\begin{subequations}\begin{align}
			h_0(\mathbf{k}) = {} &\epsilon(k_x^2 + k_y^2) + \frac{e^2\epsilon \left(A^2+\tilde{A}^2\right)}{2\hbar^2} + \frac{e^2\epsilon \tilde{A}^2}{2\hbar^2} J_{2n}(2c) + \frac{e^2\epsilon A \tilde{A}\cos\beta}{\hbar^2} \left[ (-1)^{n-1} \frac{2n}{c} J_n(c) \right] \notag \\ {} &+ \underbrace{\frac{2e\epsilon \tilde{A}}{\hbar} (-1)^n J_n(c)}_{C} \big(k_x \cos\beta + k_y \sin\beta\big) \,, \\[10pt]
			h_x(\mathbf{k}) = {} &-\alpha k_y \underbrace{- \frac{e\alpha \tilde{A}}{\hbar} (-1)^n J_n(c) \sin\beta}_{\mathcal{M}_x}\,, \\[10pt]
			h_y(\mathbf{k}) = {} &\alpha k_x \underbrace{+ \frac{e\alpha \tilde{A}}{\hbar} (-1)^n J_n(c) \cos\beta}_{\mathcal{M}_y}\,.
	\end{align}\end{subequations}
	
	\section{S3. Symmetry Analysis of the Effective Floquet Hamiltonian}
	We now analyze the derived effective Hamiltonian: $\mathcal{H}_{\rm eff}(\mathbf{k}) = h_0(\mathbf{k})\sigma_0 + h_x(\mathbf{k})\sigma_x + h_y(\mathbf{k})\sigma_y$ by defining $C = \frac{2e\epsilon \tilde{A}}{\hbar}(-1)^n J_n(c)$ and $\mathcal{M}_x, \mathcal{M}_y$.
	
	\subsection*{S3.1. Time-reversal symmetry ($\mathcal{T}$) }
	Applying $\mathcal{T}$ to the effective Hamiltonian gives 
	\begin{align}
		\mathcal{T} H_{\rm eff}(\mathbf{k}) \mathcal{T}^{-1} ={}& h_0(\mathbf{k})\sigma_0 + \alpha \big(k_x(-\sigma_y) - k_y(-\sigma_x)\big) - \delta_x(-\sigma_x) + \delta_y (-\sigma_y)= h_0(\mathbf{k})\sigma_0 + \alpha \big(-k_x\sigma_y + k_y\sigma_x\big) \notag \\{}&- \mathcal{M}_x \sigma_x - \mathcal{M}_y \sigma_y \,.
	\end{align}
	Evaluating the Hamiltonian at the reversed momentum yields\begin{align}
		H_{\rm eff}(-\mathbf{k}) ={} & \left[\epsilon k^2 - C(k_x \cos\beta + k_y \sin\beta) + \frac{e^2\epsilon }{2\hbar^2} \left(A^2+\tilde{A}^2\right)+ \frac{e^2\epsilon \tilde{A}^2}{2\hbar^2} J_{2n}(2c) + \frac{2n e^2\epsilon A \tilde{A}\cos\beta}{c \hbar^2} (-1)^{n-1} J_n(c) \right]\sigma_0 \notag \\ {} &+ \alpha\big(-k_x\sigma_y +k_y\sigma_x\big) + \mathcal{M}_x \sigma_x + \mathcal{M}_y \sigma_y \,.
	\end{align}
	Comparing the two, $\mathcal{T} H_{\rm eff}(\mathbf{k}) \mathcal{T}^{-1} \neq H_{\rm eff}(-\mathbf{k})$. Therefore, time-reversal symmetry is broken.
	
	\subsection*{S3.2. spatial inversion symmetry ($\mathcal{P}$)}
	This symmetry was already broken in the unperturbed system. The Floquet drive reinforces this breaking. The light-induced ``drift'' term $C(k_x \cos\beta + k_y \sin\beta)$ inside $h_0(\mathbf{k})$ represents a DC vector potential and the odd parity of this linear term destroys any remaining possibility of spatial inversion.
	
	\subsection*{S3.3. Two-fold rotation symmetry ($C_{2z}$)}
	This was the geometric symmetry surviving in the unperturbed Rashba 2DEG. Let us see if the light preserves it. Applying the rotation operator leads to\begin{align}
		\mathcal{U}_{C2z} \mathcal{H}_{\rm eff}(\mathbf{k}) \mathcal{U}_{C2z}^{-1} = h_0(\mathbf{k})\sigma_0 - h_x(\mathbf{k})\sigma_x - h_y(\mathbf{k})\sigma_y\,.
	\end{align}Now, we evaluate the effective Hamiltonian at the reversed momentum $-\mathbf{k}$:\begin{align}
		\mathcal{H}_{\rm eff}(-\mathbf{k}) = h_0(-\mathbf{k})\sigma_0 + h_x(-\mathbf{k})\sigma_x + h_y(-\mathbf{k})\sigma_y\,.
	\end{align}Let us isolate just the scalar identity sector $h_0$: $h_0(-\mathbf{k}) = \epsilon(-k_x)^2 + \epsilon(-k_y)^2 + C(-k_x \cos\beta - k_y \sin\beta) + \frac{e^2\epsilon \left(A^2+\tilde{A}^2\right)}{2\hbar^2} + \frac{e^2\epsilon \tilde{A}^2}{2\hbar^2} J_{2n}(2c) + \frac{2n e^2\epsilon A \tilde{A}\cos\beta}{c \hbar^2} (-1)^{n-1} J_n(c)$. Because $h_0(-\mathbf{k}) \neq h_0(\mathbf{k})$, the scalar condition fails completely. Therefore, $\mathcal{U}_{C2z} \mathcal{H}_{\rm eff}(\mathbf{k}) \mathcal{U}_{C2z}^{-1} \neq \mathcal{H}_{\rm eff}(-\mathbf{k})$. The $C_{2z}$ rotational symmetry is definitively broken. 
		
	}
\end{document}